\documentclass[preprint]{aastex}

\newcommand\gsim{\mathrel{\hbox{\rlap{\hbox{\lower4pt\hbox{$\sim$}}}\hbox{$>$}}}}

\shorttitle{Environment of USS}
\shortauthors{Bornancini et al.}

\begin{document}

\title{Clustering and light profiles of galaxies
in the environment of 20  Ultra Steep Spectrum Radio sources \altaffilmark{1}
}

\author{
Carlos G. Bornancini,
H\'ector J. Mart\'{\i}nez\altaffilmark{2},
Diego G. Lambas\altaffilmark{2}}
\affil{Grupo de Investigaciones en Astronom\'\i a Te\'orica y Experimental,
IATE, Observatorio Astron\'omico, Universidad Nacional de
C\'ordoba, Laprida 854, X5000BGR, C\'ordoba, Argentina.}
\email{bornancini@oac.uncor.edu, julian@oac.uncor.edu, dgl@oac.uncor.edu}
\author{Wim de Vries, Wil van Breugel}
\affil{Lawrence Livermore National Laboratories, L-413, 7000 East Ave,
Livermore, CA 94550, U.S.A.}
\email{wdevries@igpp.ucllnl.org, wil@igpp.ucllnl.org}
\author{Carlos De Breuck}
\affil{Institut d'Astrophysique de Paris, 98bis Boulevard Arago,
75014 Paris, France.}
\email{debreuck@iap.fr}
\and
\author{Dante Minniti}
\affil{Departamento de Astronom\'\i a y Astrof\'\i sica,
Pontificia Universidad Cat\'olica, Vicu\~na Mackenna 4860, Casilla 306
Santiago 22, Chile.}
\email{dante@astro.puc.cl}
\altaffiltext{1}
{Based on observations obtained at Cerro Tololo Inter-American Observatory,
a division of the National Optical Astronomy Observatories,
which is operated by the Association of Universities 
for Research in Astronomy, Inc.  under cooperative agreement 
with the National Science Foundation.}
\altaffiltext{2}
{Consejo Nacional de Investigaciones Cient\'\i ficas y T\'ecnicas (CONICET),
Avenida Rivadavia 1917, C1033AAJ, Buenos Aires, Argentina.}

\begin{abstract}
We have analyzed galaxy properties in the neighborhood of 20
Ultra-Steep Spectrum Radio sources (USS) taken from the WISH catalog of
De Breuck et al. (2002). Galaxies in these USS fields were identified in deep 
observations that were carried out in the $K^{\prime}-$band using the OSIRIS 
imager at the CTIO 4m telescope.
We find a statistically significant signal of clustering
around our sample of USS. The angular extension of the detected 
USS-galaxy clustering is $\theta_c\sim 20\arcsec$ corresponding to a spatial scale 
$\sim 120 h^{-1}{\rm kpc}$, assuming the sources are at 
$z \sim 1$ in a $\Omega_m=0.3$, $\Omega_{\Lambda}=0.7$ model universe.
These results are in agreement with those obtained by Best (2000) for radio 
galaxy-galaxy correlation, and Best et al. (2003) for radio-loud AGN-galaxy 
correlation. We have also analyzed the light distribution 
of the galaxies by fitting S\'ersic's law profiles.  Our results
show no significant dependence of the galaxy shape parameters
on the projected distance to the USS. 
\end{abstract}


\section{Introduction}
In current hierarchical galaxy formation scenarios such as CDM models,
the most massive galaxies at high redshifts are expected to form 
in over-dense regions corresponding to the precursors of present-day 
clusters of galaxies (e.g. White 1997). The first massive black holes 
may also grow in a similar hierarchical way than the parent galaxies where
gas infall by massive cooling flows might be the key to understand
both, galaxy and black hole formation. There is also the question 
whether massive black holes could have been formed before host galaxies
(Kauffmann \& Haehnelt 1999) and therefore the need of high $z$ census 
of radio population and galaxies. It is suggested that mergers are 
associated to powerful radio sources. Since they provide efficient 
mechanisms to trigger star formation and stimulate AGN phenomena, 
therefore it is expected some correlation between radio properties 
and star formation signatures.

At low redshift, radio sources are found in massive elliptical galaxies
and AGNs ($\sim 10^9 M_{\sun}$ accreting black holes) being
radio emission  a common feature in bright elliptical galaxies.
However, powerful radio sources are found only in bright ellipticals 
$M_B < -19$ (see for instance Lilly \& Longair 1984, Best, Longair \& 
R\"ottgering 1998). It is also found that their comoving space densities were 
much larger (100-1000 times) in the past ($z \sim 2$) than present-day 
radio galaxies.

Radio sources are good beacons for pinpointing massive elliptical galaxies
at least up to redshift $z\sim 1$. There is the well known existence 
of a very good correlation between $K-$band magnitude
and redshift for powerful radio sources (van Breugel et al. 1998,
De Breuck et al. 2001) and this appears to hold up to $z\sim 5$, despite
large $k-$corrections and morphological changes (van Breugel et al. 1999).
Therefore, radio sources may be used to find massive galaxies and 
their progenitors out to high redshift through near-IR identification.
While optical techniques have been successful in identifying `normal' 
young galaxies at high redshift, the radio and near-IR selection 
technique has the advantage that it is less biased with respect to 
the dust extinction.

Ultra-Steep Radio continuum spectrum sources correspond to sources with 
spectral index $\alpha <-1.3$ in the frequency range $352-1400$ MHz.  
These sources are less frequently identified ($< 15\%$) in POSS Plates, 
$R \le 20$, so that they are likely to be associated to foreground objects.
Radio sources with ultra steep spectrum (hereafter USS) are good candidates 
for high redshift galaxies so that identification of bright radio sources
with faint galaxies provides a convenient procedure to locate distant 
galaxies and clusters (De Breuck et al. 2001).

Previous studies suggest that the radio galaxy is not always
the brightest cluster member, although it is among the brightest galaxies.
It has been suggested by Chapman et al (2002) that radio sources
could reside in a compact environment, distinct from rich X-ray selected
clusters at similar redshifts. Best (2000) analyzing the environment
of 28 3CR radio-galaxies found a net overdensity of $K-$band galaxies 
with the mean excess counts being comparable to that expected for clusters 
of Abell class 0 richness, concluding that many powerful radio galaxies 
are located in cluster environments.
In this paper we aim to address the nature of the density enhancement
of galaxies around high redshift ultra steep spectrum radio sources
by studying the galaxy clustering around USS and the properties of the 
light distribution of galaxies in these environments.
This paper is organized as follows: Section 2 describes the sample of USS
used, the data reduction and galaxy identification. We
analyze the resulting $K'$ number counts in Section 3. In Section 4 we
study the clustering of galaxies around USS.
Section 5  deals with the morphological properties of galaxies
around USS. Finally we discuss our results in Section 6.

\section{Data}

We used 20 sources selected from the 352 MHz Westerbork In the 
Southern Hemisphere (WISH) survey restricted to have an ultra steep radio 
continuum spectrum $\alpha < -1.3$.
These USS radio sources have been selected basically on their radio properties.
The first priority targets are unresolved in the radio ({\em i.e.}
 have a point source morphology),
and the second priority targets have a small radio size ($< 30\arcsec$, 
based on the VLA
maps). All of the first and most of the second priority targets have been observed.
We have selected against large radio sources because those are most likely
foreground objects, and not at high redshift. 
This sample is aimed at increasing the number of known high redshift 
radio galaxies to allow detailed follow-up studies
of these massive galaxies and their environments in the early Universe 
(De Breuck et al. 2002). 
The galaxy sample used in this work consists 
in galaxies identified in deep images in the near infrared $K'-$band 
of our 20 USS fields, obtained in the 4$-$meter V.M. Blanco telescope 
at CTIO using the OSIRIS imager. All the images were obtained under 
optimum seeing conditions with FWHM in the range $0.5\arcsec-0.8\arcsec$.
We found clear $K'-$band counterpart of radio sources in 7 frames.
Table 1 lists the WISH designation, position, the spectral index
$\alpha_{352}^{1400}$, the $K^{\prime}$ counterpart magnitude and the 1.5 $\sigma$ 
limiting magnitude per field.

\subsection{Data acquisition and reduction}

The 20 USS fields analyzed here were observed during two runs on March 2000 and
January 2001 using the OSIRIS imager on the 4$-$meter V.M. Blanco telescope
at CTIO. The pixel scale is $0.161\arcsec/{\rm pixel}$ in a $1024\times1024$ 
pixels CCD array with an effective field of view $82\arcsec\times82\arcsec$ 
once spurious objects near boundary regions are rejected. 

We used a 16$-$point non$-$redundant dithering pattern,
which was slightly tilted to have no redundancies (i.e. none of 
the grid points land on the same row or column
number on the CCD, in this way one effectively removes row/column related
defects). Each source was observed for 1 grid, with 12 co-adds of 10s at each
grid point. This results in $16\times10\times12=1920s=32$ minutes on source. 
During the observations, tip-tilt mirror corrections were made. 
This resulted in typical FWHMs on the individual pointing frames 
($12\times10s$ co-added) of $0.5\arcsec$ to $0.7\arcsec$. The mean was 
around $0.6\arcsec$ for the nights 03/20-22/2000, and around $0.7\arcsec$ 
on the Jan 2001 run.

Both runs were photometric, and a fair number of standard stars have been 
observed on each night. The variations in the zero-points was negligible. 
Standard stars were taken from the NICMOS near-IR standard list 
(Persson et al. 1998). The magnitude calibration zero point is 22.669 mag.
Data were reduced within IRAF\footnote{Image Reduction and Analysis
Facility (IRAF), a software system distributed by the National Optical
Astronomy Observatories (NOAO)}, using the standard {\tt DIMSUM} (Stanford,
Eisenhardt, \& Dickinson 1995) near-IR reduction package. Since
the observing conditions at CTIO were good, the
resulting images are of a high quality, very flat and with uniform
backgrounds.

\subsection{Source detection and photometry}

We have used the SExtractor package version 2.1 (Bertin \& Arnouts 1996)
for photometry, object detection and galaxy-star separation in the fields.
The source extraction parameters were set so that an object
to be detected must have a flux in excess of 1.5 times the local 
background noise level over at least ten connected pixels.

SExtractor's {\tt MAG\_BEST} estimator was used to determine the magnitudes 
of the sources; this yields an estimate for the total magnitude 
using Kron's (1980) first moment algorithm, except if there is a 
nearby companion which may bias the total magnitude estimate by more than 
10\% in which case a corrected isophotal magnitude is used instead.

SExtractor provides an stellaricity index for each object, which is an 
indication of the likelihood of an object to be a galaxy or a star based 
on a neuronal network technique.  In the ideal case, a galaxy and a star have a
stellaricity index {\tt CLASS\_STAR} $=0.0$ and 1.0 respectively.
We have adopted in this work a limit in {\tt CLASS\_STAR} $< 0.8$ 
for an object to be a galaxy.

Our final catalog of galaxies in the analyzed USS frames
comprises 400 objects with $K' < 20$. $K'$ magnitudes for USS are shown
in Table 1.

\section {$K'-$band number counts}

We have computed the mean number of galaxies per unit area in the fields as 
a function of apparent $K'$ magnitude.  The results are shown in Figure 1
and quoted in Table 2.
Error bars were estimated using Poissonian errors.  In Figure 1 
we also compare our determinations with the number counts in the Subaru Deep 
Field (Totani et al. 2001), with determinations from the ESO $K'-$band 
galaxy survey (Saracco et al. 1997) and with determinations from 
NTT Deep Field $K_{s}$ galaxy counts (Saracco et al. 1999), 
assuming a mean colour $K_{s}-K'=0.2$ (Daddi et al. 2000). 
We also compare with the $K-$band surveys by Gardner, Cowie \& Wainscoat (1993)
and Best (2000), assuming a mean colour $K'-K=0.13$ (Roche et al. 1998).
All of these studies deal with field galaxies with the exception of 
Best (2000) and Best et al. (2003) which correspond to galaxies in the environment 
of radio-galaxies and galaxies in the environments of radio-loud AGN.  
We find a general agreement with these works
where the marginal excess at $K' \sim 19$ we detect could be an indication of
structure associated to the USS.

\section{Angular cross-correlation function}

In this section we analyze the clustering of galaxies around USS.
We compute the angular two point cross-correlation function $\omega(\theta)$
between the USS and the galaxies in their fields.
We have used the following estimator of the angular cross-correlation function:
\begin{equation}
\omega(\theta)=\frac{n_R}{n_G}\frac{DD(\theta)}{DR(\theta)}-1,
\end{equation}
where $n_G$ and $n_R$ are the numbers of galaxies in the sample and in a random
sample respectively, $DD(\theta)$ is the number of real pairs USS-galaxy
separated by an angular distance in the range $\theta, \theta+\delta \theta$,
and $DR(\theta)$ are the corresponding pairs when considering the random 
galaxy sample. We have done this computation for three subsamples 
of tracer galaxies defined by their apparent magnitude: $K'<18$, 
$18\leq K'\leq 19$ and $19\leq K'\leq 20$. For USS, radio position were used.  
In Figure 2 we show the resulting USS-galaxy cross-correlation functions.
We estimate correlation function error bars using the field-to-field 
variation of $DD(\theta)$. 

The angular correlation function is usually assumed to have a power-law form:
$\omega(\theta)=A\theta^{1-\gamma}$. 
If so, the observed $\omega_o(\theta)$ will follow a form
\begin{equation}
\omega_o(\theta)=A(\theta^{1-\gamma}-C),
\end{equation}
where $AC$ is known as the integral constraint and arises from the finite 
size of the field of view. The integral constraint can be computed as
\begin{equation}
AC=\frac{1}{\Omega^2}\int_{\Omega_1} \int_{\Omega_2}
\omega(\theta_{12})d\Omega_1 d\Omega_2.
\end{equation}
Using the random-random correlation, this calculation can be done numerically:
\begin{equation}
C=\frac{\sum RR(\theta)\theta^{1-\gamma}}{\sum RR(\theta)},
\end{equation}
where $RR(\theta)$ is the number of random pairs of objects with
angular distances between $\theta$ and $\theta+\delta\theta$.

For our sample geometry, and assuming $\gamma=1.8$, we find 
$C=0.0664~{\rm arcsec}^{-0.8}$. For the subsample with $18\leq K'\leq 19$ 
we find a strong correlation with an amplitude 
$A=(5.0\pm0.6)~{\rm arcsec}^{0.8}$, this is larger by a factor $\sim 2$ 
than the galaxy-galaxy correlation amplitude for $K<19$ obtained by Best (2000). 
On the other hand the $K'<18$ subsample
has a negligible cross-correlation amplitude 
$A=(0.44\pm0.64)~{\rm arcsec}^{0.8}$. 
The strong cross-correlation signal between USS and galaxies
with $18\leq K'\leq 19$ implies that they are physically associated
on scales smaller than $\sim 20"$. This angular scale corresponds to 
$\sim 120 h^{-1}{\rm kpc}$, assuming the sources at $z\sim 1 $
(average redshift estimated using the Hubble $K-z$ diagram taken from 
De Breuck el al. 2002), and a flat cosmological model with density 
parameters $\Omega_0=0.3$ and $\Omega_{\Lambda}=0.7$ and a Hubble's constant
$H_0=100 ~h ~{\rm kms}^{-1}~{\rm Mpc}^{-1}$. This is not the case for 
galaxies brighter than $K'=18$, where we detect no significant correlation 
signal suggesting they are mostly foreground objects.

We have tested for the possibility of a biased tendency of detecting 
faint objects near the central positions of the frames where USS are located.
We have computed USS-galaxy cross correlations for faint galaxies, 
$19 < K' <20$ and we find no significant correlation signal which indicates 
the lack of systematic effects in our analysis (see Figure 2).

Recall that when analyzing number counts we have found a marginal excess of
$K'\sim 19$ with respect to previous determinations.
Both, the cross-correlation analysis and the number counts
lead us to support the idea that USS are related to high-$z$ radio galaxies
located in protocluster environments.

One might question whether our galaxy sample is deep enough to provide
a useful insight on the galaxy environment around these USS (estimated
to be at $z\sim 1$, see above). First, we notice that the 1.5 $\sigma$ detection limits
for the fields analyzed are significantly fainter than our cutoff at $K'=20$. Also,
one may consider other works that 
deal with $K-$band photometry of galaxies in clusters at 
redshifts $z \gsim 1$ as for instance Stanford et al. (2002),
 Blanton et al. (2003), 
and Stanford et al. (1997). In these studies, cluster member galaxies 
with measured spectroscopic redshifts have 
$K-$band magnitudes within the range of our analysis, 
indicating that our galaxy sample depth is appropriate for our study.

\section {Light distribution profiles}
To deepen our understanding on the nature of galaxies in the overdensities
around USS, we provide a measure of galaxy morphology in the USS fields by 
analyzing the light distribution of galaxies in the images
using IRAF {\tt STSDAS ELLIPSE} package (Jedrzejewski 1987).

Since S\'ersic's law (S\'ersic 1968) has proved to be useful in the 
characterization of galaxy luminosity profiles, we use it to fit the light 
distribution of tracer galaxies in our USS fields.
S\'ersic's law can be written as:
\begin{equation}
\Sigma(r)=\Sigma(r_e)\exp
\left(-b_n\left[\left(\frac{r}{r_e}\right)^{1/n}-1\right]\right)
\end{equation}
where $\Sigma(r)$ is the surface brightness at radius $r$. The parameter 
$b_n$ is set equal to $1.9992n-0.3271$, so that $r_e$ remains the projected 
radius enclosing half of the galaxy's light. We recall that $n \sim 1$ 
corresponds to an exponential  profile while $n \sim 4$ corresponds to
a de Vaucouleurs profile, characteristic of disks and early-type objects, 
respectively.

As analyzed in Section 4, galaxies with $18<K'<19$ show a significant 
cross-correlation signal so that a large fraction of them are likely to be 
associated to USS. We have studied the light 
profiles and the corresponding S\'ersic's law fits. Due to the low signal 
to noise of many images we restricted our analysis to those profiles resolved 
with significant light well beyond the radius of the seeing disk.
This results in a subsample of 31 galaxies, {\em i.e.} $\sim 25\%$ of the 
objects in the considered magnitude range. Four examples of our fitting 
procedure are displayed in Figure 3.

The interpretation of the observed shape parameter $n$ has been
assessed using a set of simulated galaxies, which is discussed
in the next subsection.

\subsection{Simulating model galaxies to account for PSF effects}

In order to establish the relation between observed and actual values 
of the shape index $n$, we have tested our results on a set of simulated 
galaxies which are affected by PSF convolution.

We have chosen a set of 15 galaxies in our sample with effective radii 
in the range $0.2\arcsec-1.5\arcsec$, different magnitudes 
(in the range of $18<K'<19$) 
and observed S\'ersic's shape parameters $n$. We created model galaxies 
similar to these and convolved them with the observed PSF to 
match the resolution of the data. Model galaxy images were built using 
the IRAF {\tt bmodel} task and convolved  using  IRAF {\tt gauss} task 
with a Gaussian function representing the PSF effects. The PSF FWHM in the 
simulation was set to $0.6\arcsec$, consistent with the average of our observations.
Then, we fit S\'ersic law profiles to the original and to the convolved 
model galaxies to determine the differences between input and observed 
values due to the effects of the PSF. In order to simulate realistic 
observed conditions in our simulations, we have included the contribution
of the background noise present in our images. The background level is 
measured on blank sky regions close to the sources. We have estimated the 
uncertainty in the shape parameter values $n$ of the convolved galaxy 
images using the typical background noise fluctuation (1 $\sigma$ level).

According to the results shown in Figure 4, it can be seen that
 objects with observed index
$n<1.3$ correspond to galaxies with actual shape
index values typical of late types.
This result is similar to that obtained by
Moth \& Elston (2002) in their analysis of Hubble Deep Field galaxies.
For $n>1.3$ the models show consistency with early type objects,
that is intrinsic values $n\ge3$.
Therefore, we adopt this threshold in observed $n$ values to discriminate
between late and early type galaxy morphologies.

We find that it is critical to take into account for the PSF effects when 
drawing conclusions from luminosity profiles since at small galaxy sizes 
and faint magnitudes, an $n=4$ profile can be substantially smoothed out.
We find a tendency for high $n$ to be underestimated, furthermore,
the higher the original $n$ value, the more it will be affected by the PSF
(Moth \& Elston 2002).

\subsection{Light profiles properties}

In Figure 5 we show the distribution of S\'ersic's shape parameter $n$.
The number of early type galaxies (14) almost equals that of late
type galaxies (17), according to our criterion. The scatter-plot of $n$ vs. 
the angular distance to the radio source is shown in Figure 6,
where uncertainties in shape parameters were estimated using $1\sigma$ background
noise fluctuation as mentioned before.
 
The Spearman rank correlation coefficient for this relation is $r=-0.083$, 
indicating a lack of significant correlation between $n$ and the projected 
distance to the radio source. It should be noticed that the results shown 
in this figure are not likely to be biased by a selection criterion that 
could affect the conclusions. In fact, from the number of galaxies with 
measured light profiles as a function of the angular distance from the 
USS and the cross-correlation analysis of section 4, we infer that this 
subsample of 25 \% of the galaxies is consistent with being extracted at 
random from the total sample in the USS fields. 

Lubin, Oke \& Postman (2002) found a trend of 
decreasing early-type fraction of galaxies in clusters with redshift, 
$f_e\sim 0.5$ at $z\sim 1$. Although with poor number statistics,
our results also suggest a similar low 
fraction of early types associated to USS overdensities.

\section{Conclusions}

We have identified galaxies in deep $K^{\prime}$-band CCD frames centered 
in 20 Ultra Steep Spectrum radio sources selected from the WISH survey. 
These observations obtained with the OSIRIS 
imager at CTIO 4m telescope were carried out with excellent seeing conditions 
$\sim 0.6\arcsec$. We have performed statistical analysis of 
non-stellar objects in these frames 
in order to shed light on the properties of galaxies in the neighborhood of USS.

We find a strong correlation signal of galaxies with $18<K^{\prime}<19$ 
around the USS. The angular extension of the detected USS-galaxy clustering 
is small ($\sim 20\arcsec$), which would correspond to $120 h^{-1}{\rm kpc}$
(assuming the sources at $z \sim 1$ in the adopted cosmology).

The evidence from our cross-correlation analysis and number counts
suggests that USS are located in protocluster environments
at high redshifts.

Light distribution profiles of galaxies in the 
frames using S\'ersic's law fits indicate a lack of 
strong dependence of the shape parameter $n$ with the projected distance 
to the radio source.

\acknowledgements

This work was partially supported by the Consejo Nacional de Investigaciones 
Cient\'{\i}ficas y T\'ecnicas (CONICET), Agencia de Promoci\'on de Ciencia y 
Tecnolog\'{\i}a,  Fundaci\'on Antorchas, Secretaria de Ciencia y
T\'ecnica de la Universidad Nacional de C\'ordoba (SECyT) 
and Agencia C\'ordoba Ciencia. 
We acknowledge support from the SETCIP/CONICYT joint grant CH-PA/01-U01.
Carlos De Breuck is supported by a Marie Curie Fellowship of the European Community
program `Improving Human Research Potential and the Socio-Economic
Knowledge Base' under contract number HPMF-CT-2000-00721.
The work by Wim de Vries, Carlos de Breuck, and Wil van Breugel
was performed under the auspices of the U.S Department of Energy,
National Nuclear Security Administration by the University of California,
Lawrence Livermore National Laboratory under contract N$^{\circ}$ W-7405-Eng-48.
Dante Minniti is supported by FONDAP Center for Astrophysics 15010003.

\clearpage
\begin{figure}
\plotone{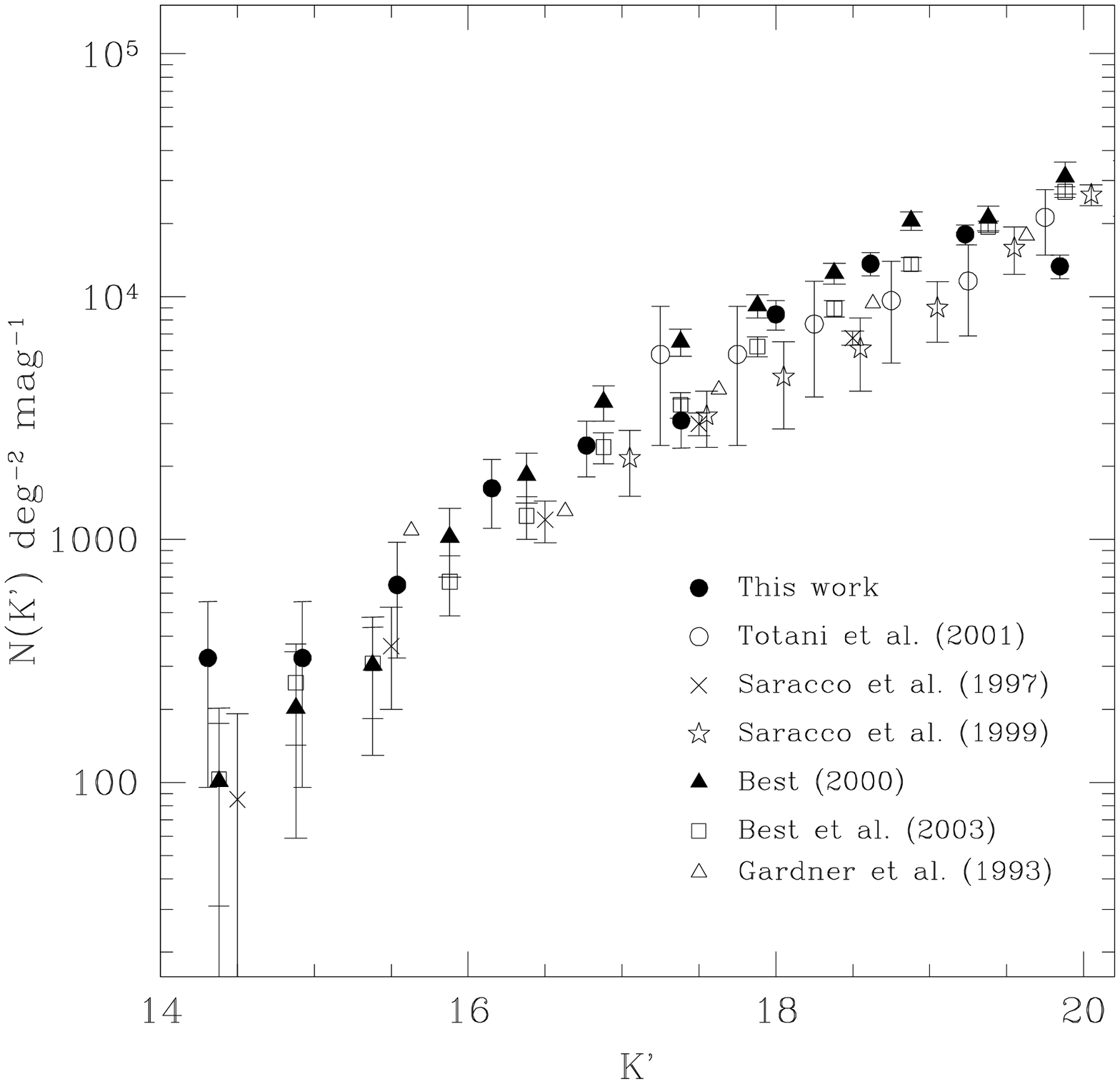}
\caption{
Filled circles: $K^{\prime}-$band galaxy number counts per magnitude
per square arcminute for our 20 USS fields. Open circles are the number 
counts from faint galaxies in the Subaru Deep Field (Totani et al. 2001).
Crosses represent data from the ESO $K^{\prime}-$band galaxy survey
(Saracco et al. 1997). Stars represent data from NTT Deep Field 
(Saracco et al. 1999). Filled triangles are the number counts from 
galaxies in the environments of 3CR galaxies (Best 2000). 
Open squares represent data of radio-loud AGN environment (Best et al. 2003).
Open triangles represent data from Gardner et al. (1993).}
\end{figure}

\begin{figure}
\plotone{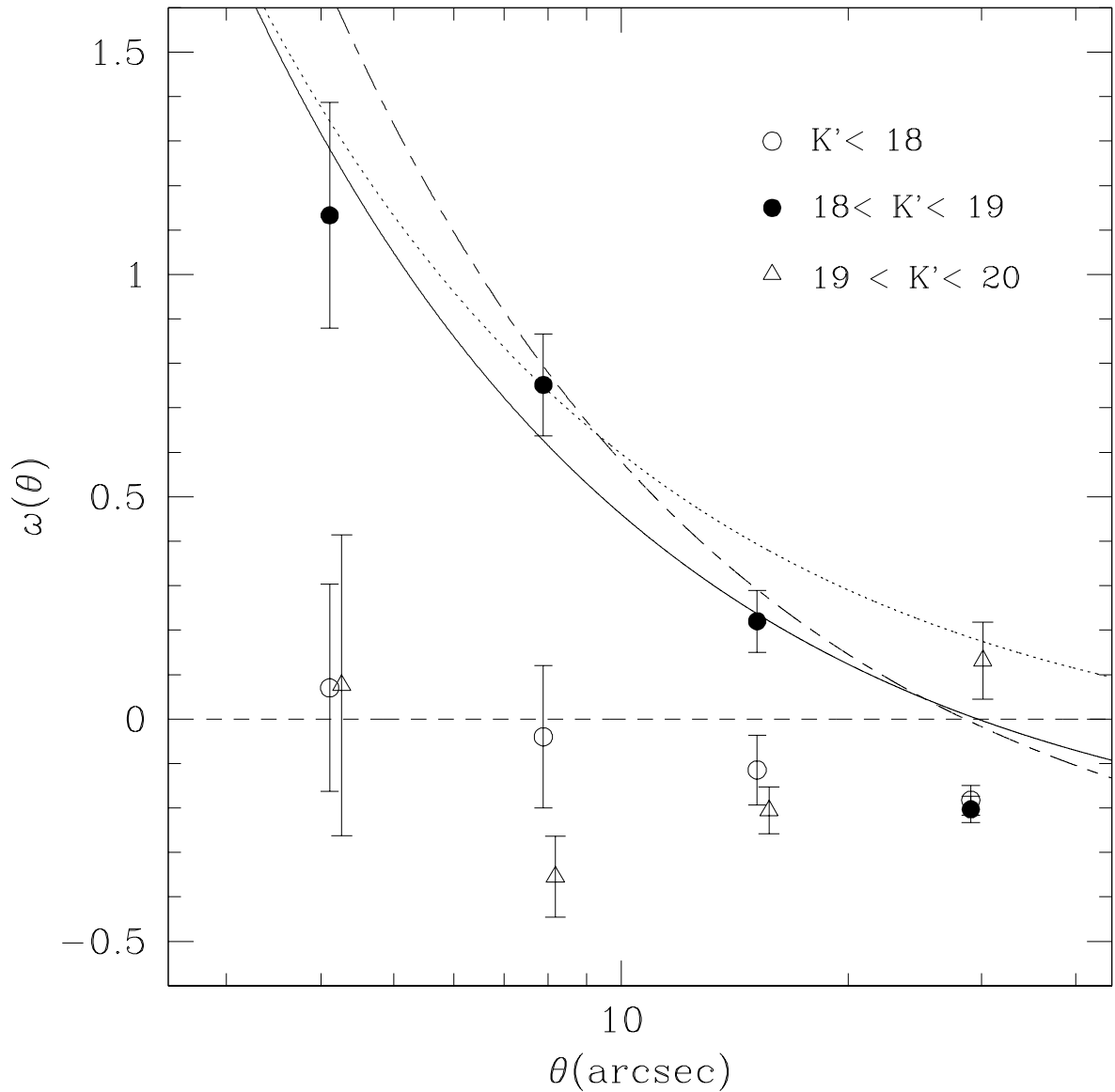}
\caption{The angular USS-galaxy cross-correlation function
for three subsamples of tracer galaxies defined by $K'<18$ (open circles),
$18\leq K'\leq 19$ (filled circles) and $19\leq K'\leq 20$ (open triangles). 
Solid line is the best power law fit to cross-correlation for
tracer galaxies with $18\leq K'\leq 19$. Dotted line corresponds 
to $K<19$ radio galaxy-galaxy correlation function taken from Best (2000) and
short-long dashed line is the radio-loud AGN-galaxy correlation 
function taken from Best et al. (2003).}
\end{figure}

\begin{figure}
\includegraphics[width=8.7cm,height=8.7cm]{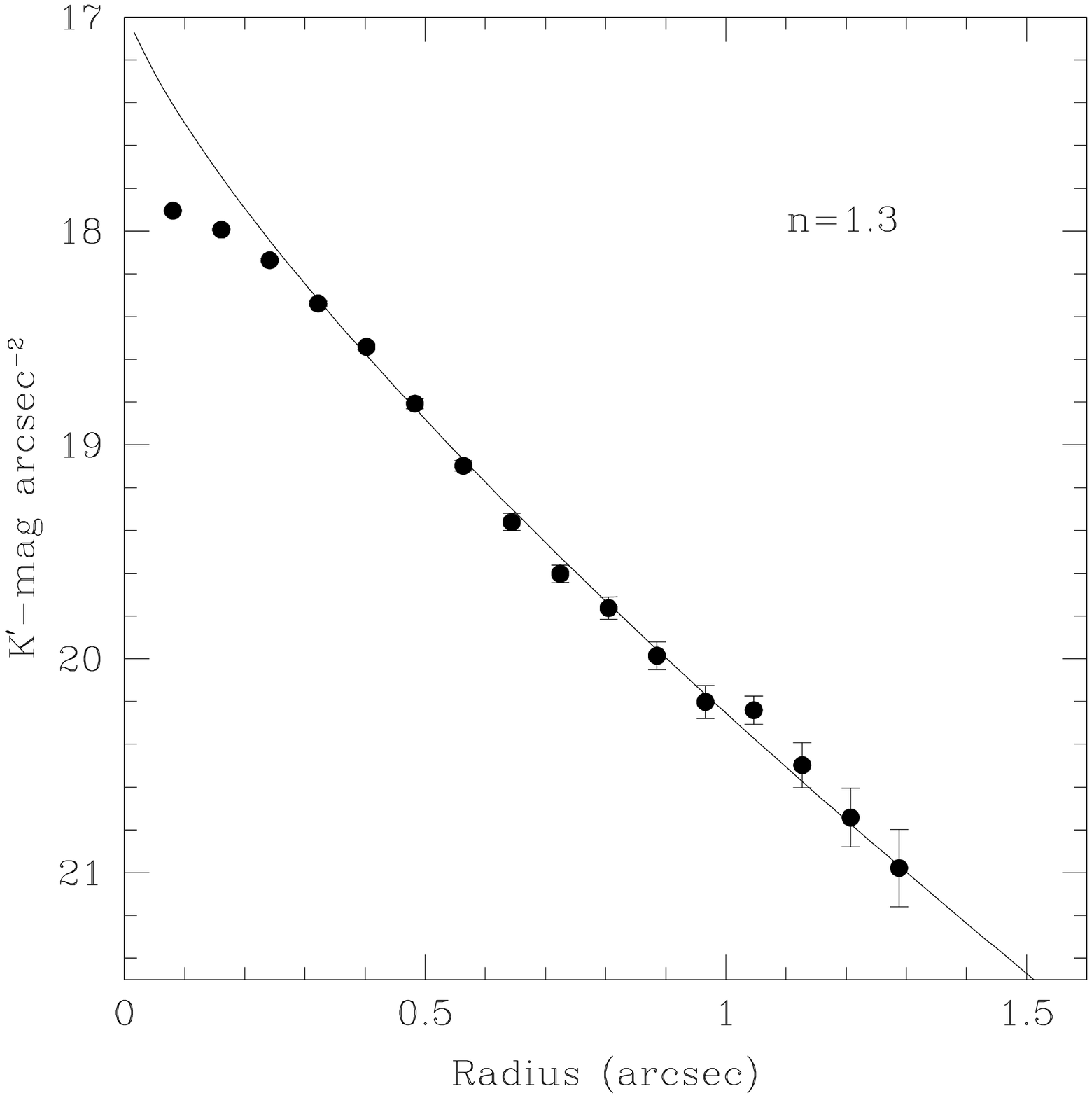}
\includegraphics[width=8.7cm,height=8.7cm]{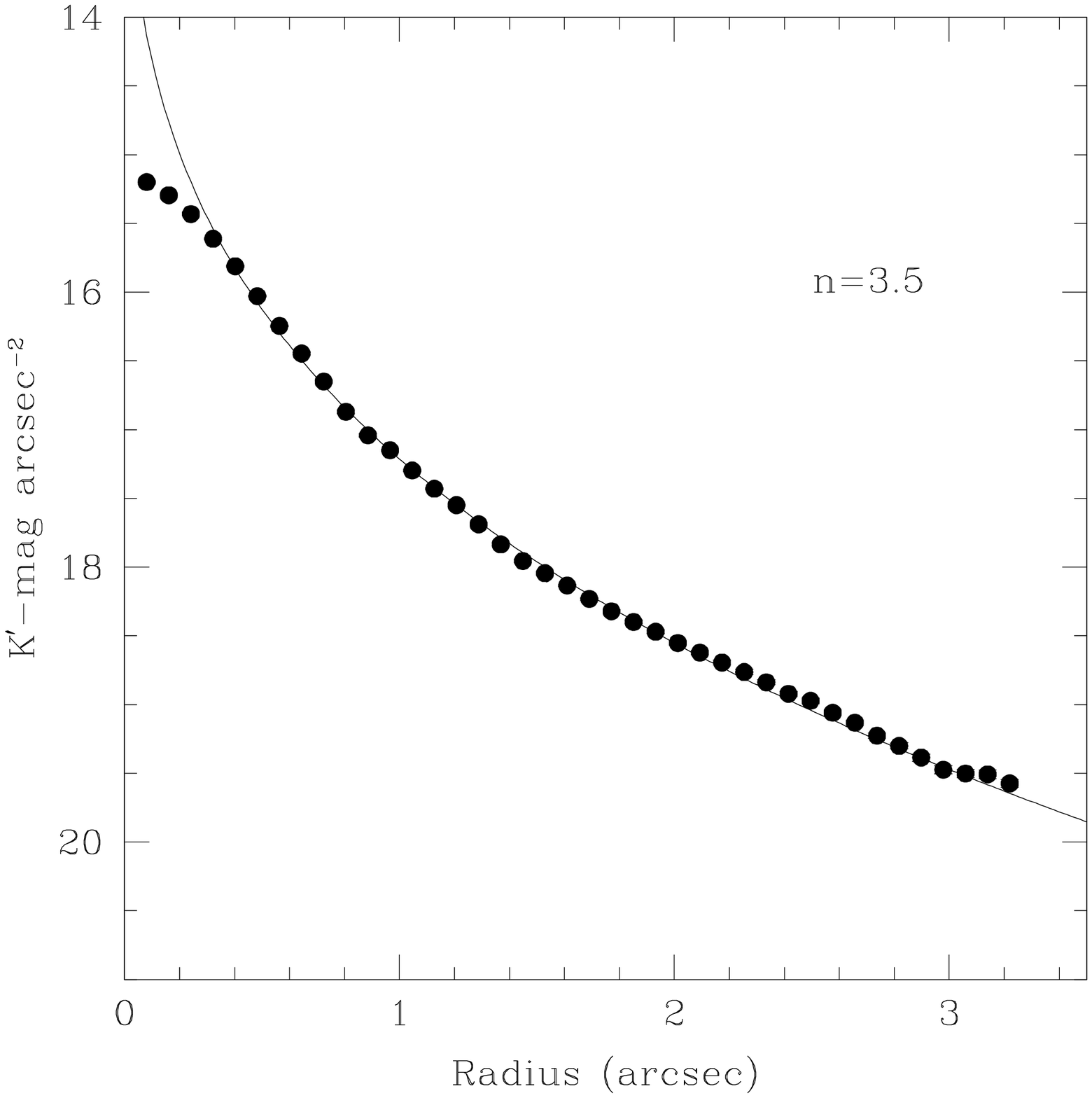}
\includegraphics[width=8.7cm,height=8.7cm]{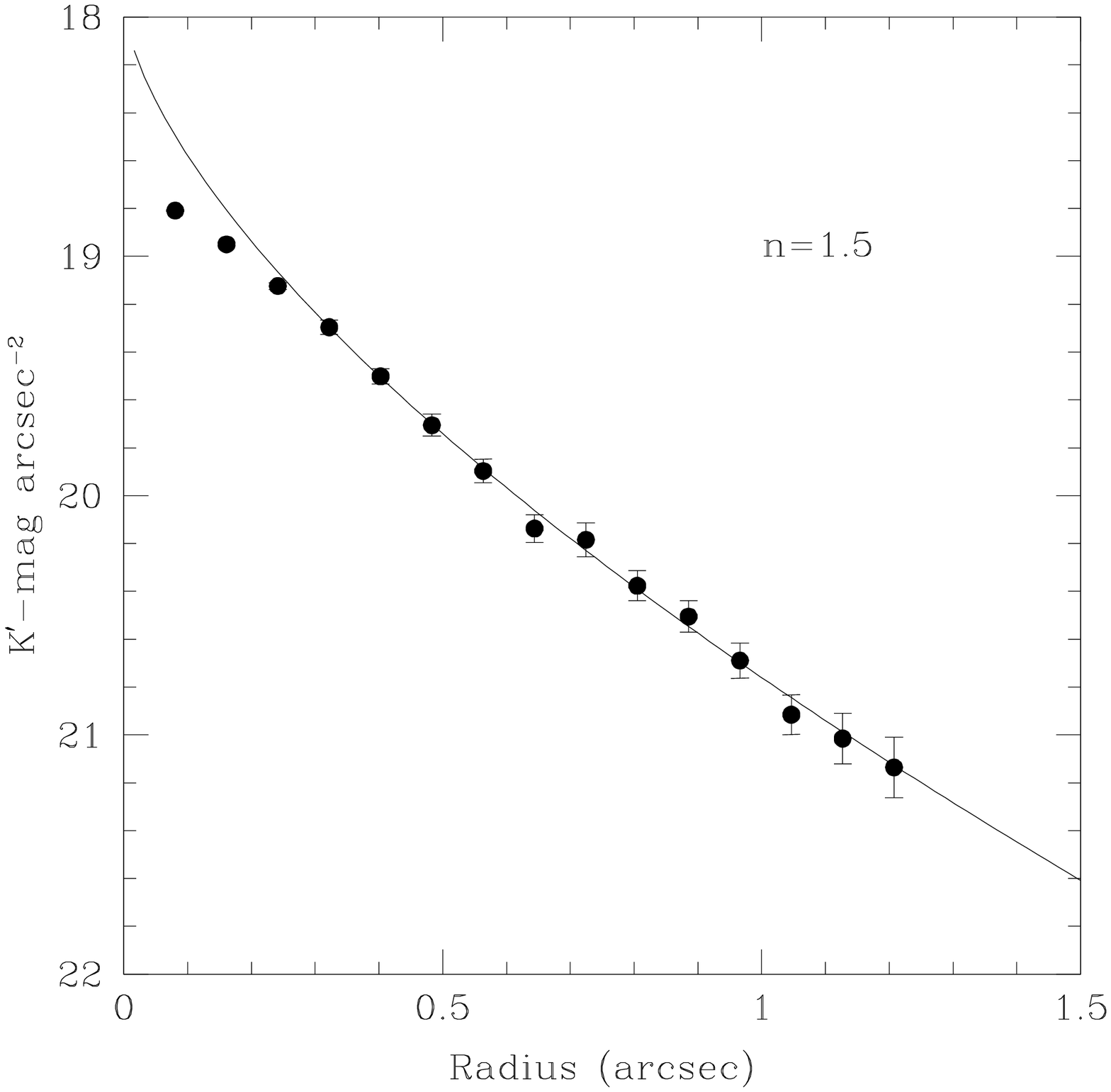}
\includegraphics[width=8.7cm,height=8.7cm]{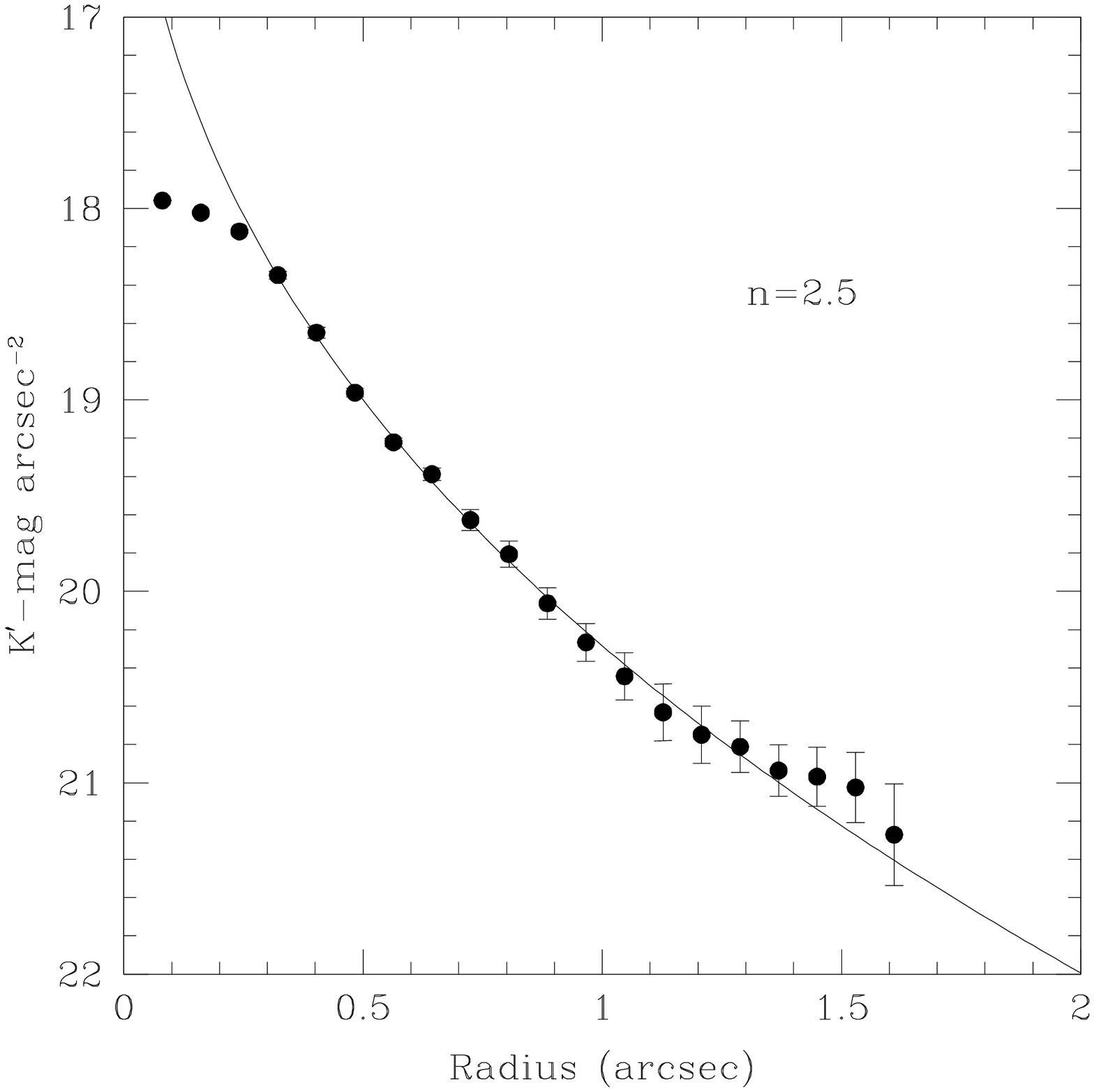}
\caption{Luminosity profiles for 4 galaxies in the sample.
The solid lines are best-fitting S\'ersic profiles whose parameters $n$ 
are given in the panels. Upper left, a typical disk, $n=1.3$, 
galaxy profile near the radio source WN J1603-1500. 
Upper right, early type-like light profile, $n=3.5$, in the field of
the radio source WN J0612-1536. Bottom left, a $n=1.5$ profile, $K'$ 
counterpart of the radio source WN J1047-1836.
Bottom right, $n=2.5$, $K'$ counterpart of the radio source WN J1331-1947. }
\end{figure}

\begin{figure}
\plotone{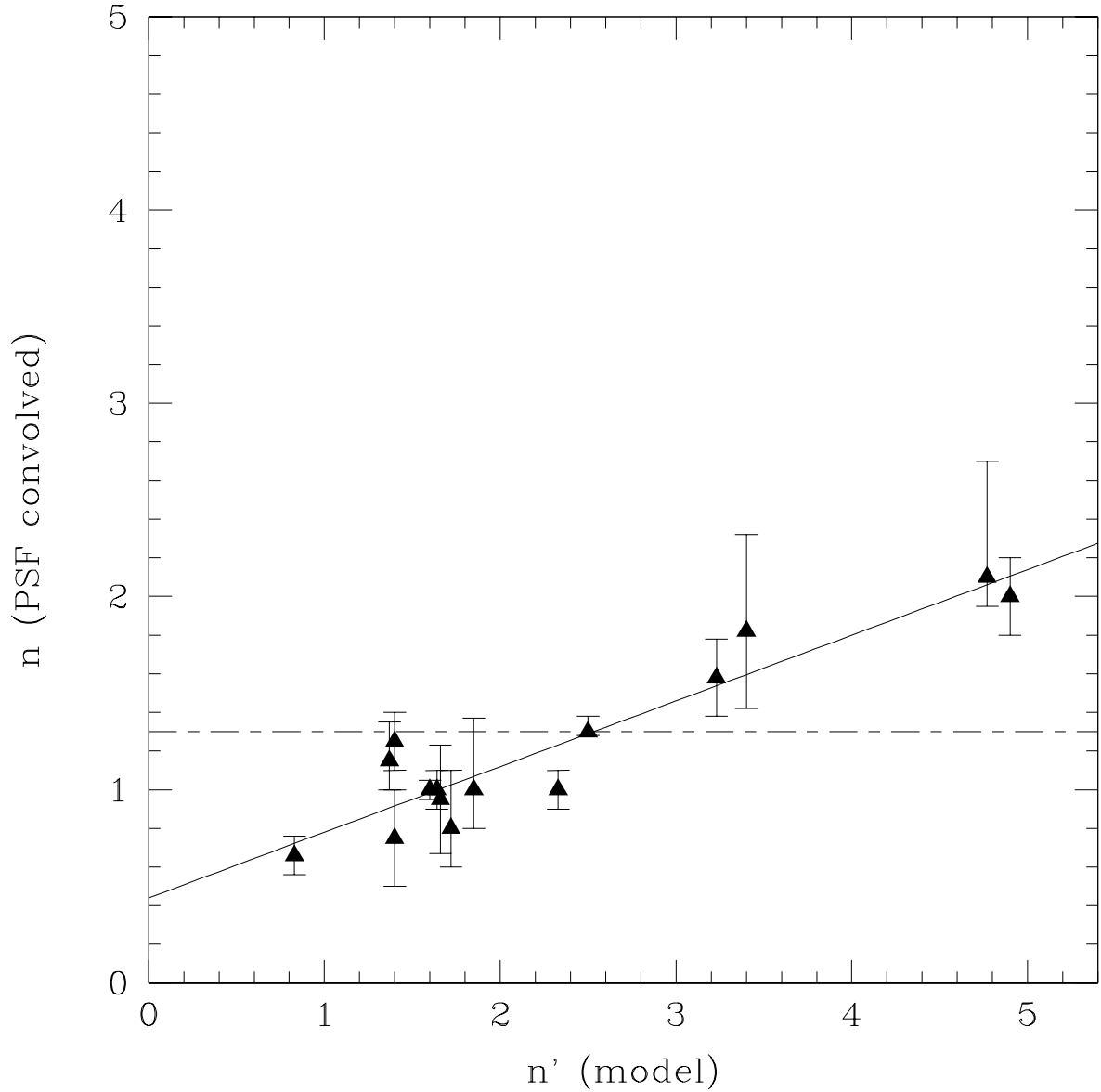}
\caption{Model and model-convolved S\'ersic's shape parameter $n$ for 
simulated galaxies.  Uncertainties in the convolved images shape parameter 
values $n$ were estimated by changing the background noise in 1 
standard deviation.}
\end{figure}

\begin{figure}
\plotone{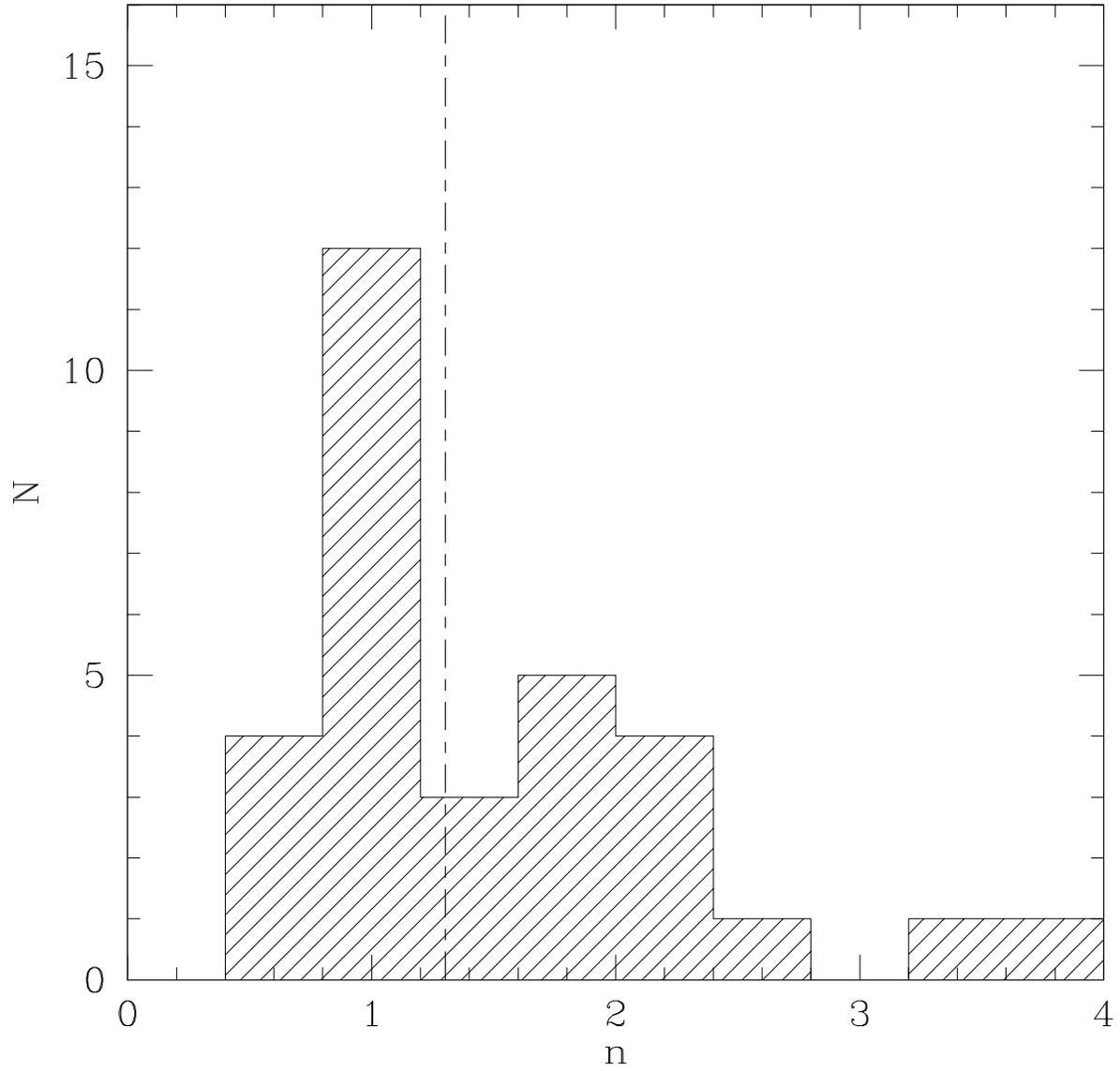}
\caption{Distribution of observed S\'ersic's shape parameter for objects 
with $18<K^{\prime}<19$. The long-short dashed line shows the limit for 
galaxy classification after taking into account PSF effects based on galaxy 
models discussed in the text.}
\end{figure}

\begin{figure}
\plotone{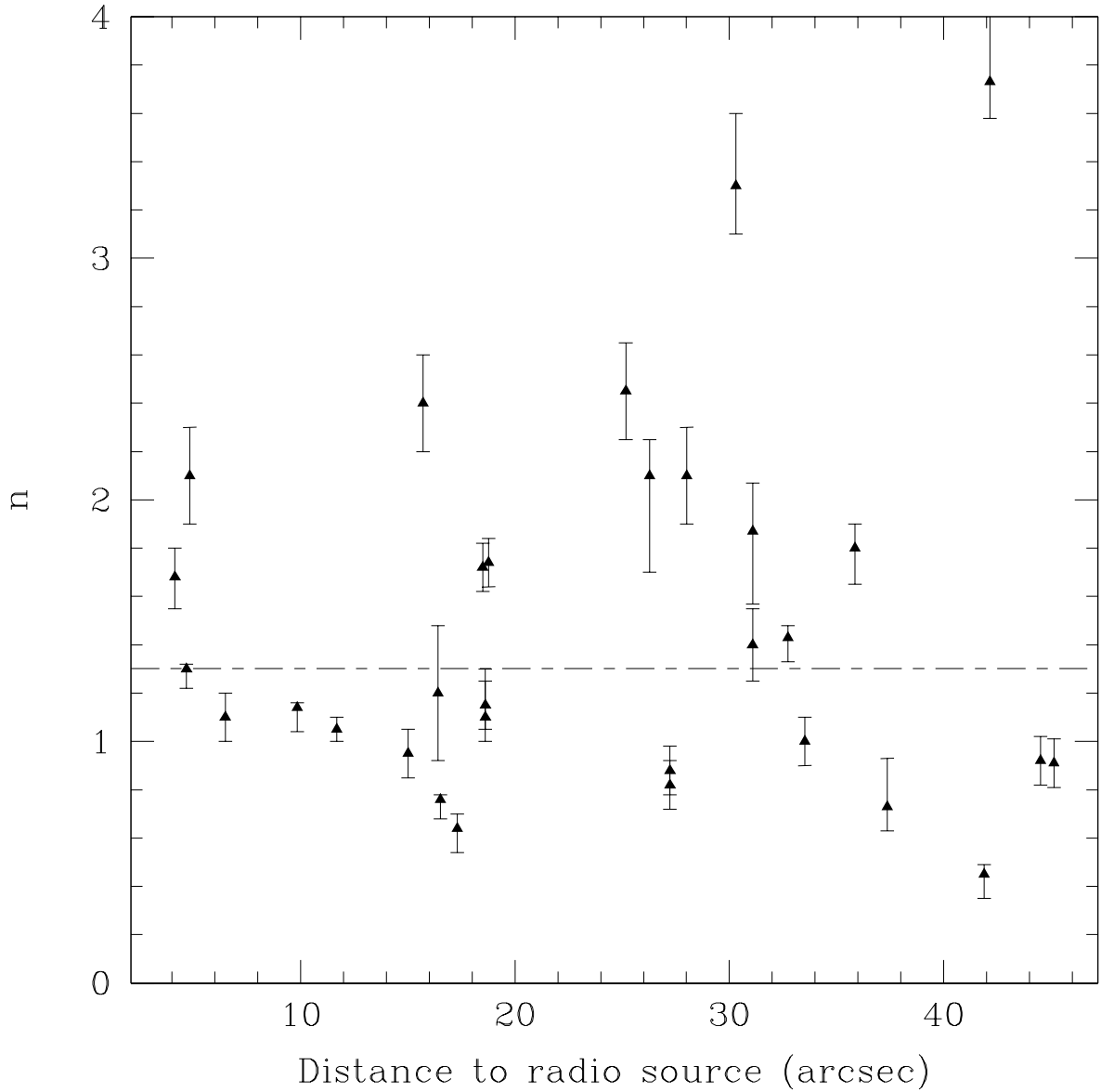}
\caption{Observed S\'ersic's shape parameter $n$ vs. projected distance
to radio source, for galaxies with $18<K^{\prime}<19$. The long-short 
dashed line shows the limit for galaxy classification as explained in text.}

\end{figure}

\clearpage
\begin{deluxetable}{cccccc}
\tablecaption{USS sample characteristics}
\label{uss}
\tablewidth{0pt}
\tablehead{
\colhead{Name} & 
\colhead{$\alpha_{J2000}$} & 
\colhead{$\delta_{J2000}$} & 
\colhead{$\alpha_{352}^{1400}$} &  
\colhead{$K^{\prime}$mag} &
\colhead{Limiting} \\
 & 
\colhead{$^h\;\; ^m\;\;\; ^s\;\;\,$} & 
\colhead{\degr$\;\;\;$ \arcmin$\;\;\;$ \arcsec$\;$} & 
 & 
 & 
\colhead{magnitude}\\
& & & & & \colhead{1.5 $\sigma$} 
}
\startdata
WN~J0526$-$1830 & 05 26 24.60 & $-$18 30 40.1 &  $-$1.39$\pm$0.06 & &        	 20.5\\
WN~J0612$-$1536 & 06 12 38.23 & $-$15 36 47.2 &  $-$1.47$\pm$0.06 & & 	         20.2\\
WN~J0910$-$2228 & 09 10 34.15 & $-$22 28 43.3 &  $-$1.57$\pm$0.04 & 19.02$\pm$0.15 & 22.1\\
WN~J0912$-$1655 & 09 12 57.24 & $-$16 55 54.8 &  $-$1.58$\pm$0.06 & 17.68$\pm$0.07  & 20.0\\
WN~J1047$-$1836 & 10 47 15.51 & $-$18 36 31.1 &  $-$1.46$\pm$0.05 & 18.66$\pm$0.09 &  20.0\\
WN~J1052$-$1812 & 10 52 00.82 & $-$18 12 32.3 &  $-$1.52$\pm$0.05 & 17.05$\pm$0.03 &	 21.1\\
WN~J1101$-$2134 & 11 01 53.63 & $-$21 34 28.5 &  $-$1.67$\pm$0.07 & 		&	 21.1\\
WN~J1109$-$1917 & 11 09 49.93 & $-$19 17 53.7 &  $-$1.38$\pm$0.04& 	&        20.0\\
WN~J1138$-$1324 & 11 38 05.42 & $-$13 24 23.5 &  $-$1.53$\pm$0.08 & 	&        	 20.5\\
WN~J1150$-$1317 & 11 50 09.59 & $-$13 17 53.9 &  $-$1.37$\pm$0.04  & 19.18$\pm$0.13 & 20.3\\
WN~J1222$-$2129 & 12 22 48.22 & $-$21 29 10.0 &  $-$1.42$\pm$0.06 & 	& 	 20.0\\
WN~J1255$-$1913 & 12 55 52.66 & $-$19 13 00.6 &  $-$1.67$\pm$0.06 &    	&        20.0\\
WN~J1331$-$1947 & 13 31 47.18 & $-$19 47 26.5 &  $-$1.40$\pm$0.04 & 18.01$\pm$0.07 & 20.0\\
WN~J1450$-$1525 & 14 50 42.63 & $-$15 25 45.2 &  $-$1.42$\pm$0.11 &     & 		 21.5\\
WN~J1516$-$2110 & 15 16 42.32 & $-$21 10 27.4 &  $-$1.38$\pm$0.04 & 	&        20.0\\
WN~J1518$-$1225 & 15 18 43.43 & $-$12 25 35.6 &  $-$1.67$\pm$0.06 & 		&        20.3\\
WN~J1557$-$1349 & 15 57 41.72 & $-$13 49 54.8 &  $-$1.39$\pm$0.06 & 18.4$\pm$0.1& 	 20.0\\
WN~J1603$-$1500 & 16 03 04.78 & $-$15 00 53.8 &  $-$1.44$\pm$0.05 &  		&	 20.5\\
WN~J1637$-$1931 & 16 37 44.85 & $-$19 31 22.9 &  $-$1.60$\pm$0.04 &  	&	 20.5\\
WN~J2002$-$1842 & 20 02 56.00 & $-$18 42 47.8 &  $-$1.42$\pm$0.06 &  		&	 20.0\\
\enddata
\end{deluxetable}

\clearpage
\begin{deluxetable}{cccc}
\tablecaption{
Galaxy counts as a function of $K^{\prime}$-band magnitude. The columns 
give the raw counts $n_{\rm raw}$,counts per square degree magnitude $N_{g}$ 
and its uncertainty $\delta N_{g}$ estimated using Poissonian errors.}
\tablewidth{0pt}
\tablehead{
\colhead{$K^{\prime}$} & 
\colhead{$n_{\rm raw}$} & 
\colhead{$N_{g}$}  & 
\colhead{$\delta N_{g}$} \\
 & & 
\colhead{$({\rm deg}^{-2}{\rm mag}^{-1})$} & 
\colhead{$({\rm deg}^{-2}{\rm mag}^{-1})$} \\
}
\startdata
$14.0 - 14.6$   &2   &324   &229  \\
$14.6 - 15.2$   &2   &324   &229  \\
$15.2 - 15.8$   &4   &655   &327  \\
$15.8 - 16.4$   &11  &1787  &538  \\
$16.4 - 17.0$   &15  &2437  &629  \\
$17.0 - 17.6$   &19  &3087  &708  \\
$17.6 - 18.2$   &52  &8449  &1171  \\
$18.2 - 18.8$   &84  &13649 &1484  \\
$18.8 - 19.4$   &111 &18036 &1711  \\
$19.4 - 20.0$   &82  &13324 &1471  \\
\enddata
\end{deluxetable}

\clearpage

\end{document}